\documentclass[notitlepage,superscriptaddress,aps,prd,nofootinbib]{revtex4-1}
\usepackage[T1]{fontenc}
\usepackage[latin9]{inputenc}
\usepackage{amsmath,amssymb}
\usepackage{xcolor}
\usepackage{graphicx}
\begin{document}
\title{Ricci-inverse gravity: a novel alternative gravity,\\
its flaws, and how to cure them}
\author{Luca Amendola}
\affiliation{Institut f\"{u}r Theoretische Physik, Ruprecht-Karls-Universit\"{a}t Heidelberg, Philosophenweg 16, D-69120 Heidelberg, Germany}
\author{ Leonardo Giani}
\affiliation{N\'ucleo Cosmo-ufes and PPGCosmo, Universidade Federal do Esp\'irito Santo,\\ Avenida Fernando Ferrari 514, 29075-910 Vit\'oria, Esp\'irito Santo, Brazil}
\affiliation{Institut f\"{u}r Theoretische Physik, Ruprecht-Karls-Universit\"{a}t Heidelberg, Philosophenweg 16, D-69120 Heidelberg, Germany}

\author{ Giorgio  Laverda}
\affiliation{Institut f\"{u}r Theoretische Physik, Ruprecht-Karls-Universit\"{a}t Heidelberg, Philosophenweg 16, D-69120 Heidelberg, Germany}
\begin{abstract}
    We introduce a novel theory of gravity based on the inverse of the Ricci tensor, that we call the anticurvature tensor. We derive the general equations of motion for any Lagrangian function of the curvature and anticurvature scalars. We then demonstrate a no-go theorem: no Lagrangian that contains terms linear in any positive or negative power of the anticurvature scalar can drive an evolution from deceleration to acceleration, as required by observations. This effectively rules out many  realizations of this  theory, as we illustrate in detail in a particular case. Finally, we speculate on how to circumvent the no-go theorem.
\end{abstract}

\maketitle

\section{Introduction}

Alternative theories of gravity have been investigated ever since the first Einsteinian formulation more than one hundred years ago. The motivations have been varied: from the inclusion of electrodynamics in Kaluza-Klein extra-dimensional metrics \cite{Kaluza:1921tu,Klein:1926tv}, to the "large number hypothesis" of Dirac \cite{Dirac:1937ti,Dirac1938}, to the search for an "affine theory" by Eddington \cite{1923mtr..book.....E} and Schroedinger \cite{schrodinger}. More recently, the main motivation came from the puzzling observation of the cosmic acceleration \cite{Riess:1998cb,Perlmutter:1998np}, and the difficulty of explaining it with a unnaturally fine-tuned value of the cosmological constant \cite{Weinberg,Padilla:2015aaa}. In addition, the so-called tensions of the standard $\Lambda$CDM \cite{Bernal:2016gxb,Battye:2014qga} have prompted the creativity of theorists to find explanations in modification of gravity.

One unpleasant characteristic of most alternative formulations is the introduction of new dimensional constants. In order for the models to have an observable effect at the present epoch, this constant needs to be of the order of powers of $H_0$, which immediately brings back the fine-tuning problem. 
Another problem is the introduction of new fields, scalar or vectors,  that have no obvious relation to the curvature tensor, marking a dramatic departure from Einstein's purely geometric formulation. 
These issues are of course not necessarily  showstoppers, but  neither are they  arguments in favour of the modification of Einstein's gravity. At the very least, however, they motivate the search for alternative gravity theories based entirely on the curvature tensor and adding no new dimensional constants. This paper is devoted to a novel gravity theory that has both properties.

As far as we know, there is only one case studied in the literature which satisfy these properties, namely the non-local model of Deser and Woodard (DW) \cite{Deser:2007jk,Woodard:2014iga,Deser:2019lmm}. One can of course easily write down combinations $f(K)R$ of the curvature tensor with the same dimensions as $R$, e.g. with $K\equiv R^{-2} R^{\mu\nu}R_{\mu\nu}$, such that   a modification of gravity without new mass scales arises. However, these terms are very complicated and consequently have not been studied in any detail so far.
The DW term is instead in principle quite simple, being based on the additional term $R \Box ^{-1} R$. It has been shown that it can produce an accelerated phase and growth of perturbations in agreement with observations \cite{Park:2012cp,Dodelson:2013sma,Nersisyan:2017mgj,Amendola:2019fhc}
without fine-tuning of parameters, at the price however of a complicate function $f(\Box ^{-1} R)$.

Motivated by the previous considerations,  we introduce in this paper a novel way of building an alternative gravity theory. We define the  tensor $A^{\mu\nu}$  as the inverse of $R_{\mu\nu}$, i.e. such that
\begin{equation}
    A^{\mu\nu} R_{\nu\sigma}=\delta^\mu_\sigma \; .
    \label{eq:delta}
    \end{equation}
    We call $A^{\mu\nu}$ the {\it anticurvature tensor}.
 One can then build the anticurvature scalar $A=g_{\mu\nu}A^{\mu\nu}$ (of course $A\not = R^{-1}$).
 With $A$, it is easy to construct relatively simple terms with the same dimensions as $R$, for instance $A^{-1}$ and $R^2 A$ or in general any $f(RA)$R.

Having introduced a theory of gravity based on the inverse Ricci tensor, we immediately proceed to prove a no-go theorem: any Lagrangian $f(R,A)$ function of the curvature and anticurvature scalars that contains terms proportional to $A^n$, with any positive or negative $n$, cannot contain both a decelerated and an acceleration cosmic expansion. As a consequence, they are ruled out as dark energy candidates. This powerful theorem is actually quite simple to demonstrate, since it turns out that both $A$ and $A^{-1}$ are singular at some epoch that is intermediate between deceleration and acceleration.

A theory based on the anticurvature tensor is a type of fourth-order gravity, similar to models built with terms like $R^{\mu\nu}R_{\mu\nu}$ and $R^{\alpha\beta\gamma\delta}R_{\alpha\beta\gamma\delta}$, extensively studied in the past \cite{Carroll:2004de,Allemandi:2004wn,Li:2007xw,CAPOZZIELLO2011167}. 
The inverse of the Ricci tensor can be expanded into the ratio of a third-order polynomial in $R^{\mu\nu}$ divided by a fourth-order polynomial, that is, an extremely non-linear function of the Ricci tensor\footnote{We thank Ignacy Sawicki for pointing  this out to us.}, analogously to the Einstein-Born-Infeld theory \cite{1998CQGra..15L..35D}.
A Lagrangian which is a generic function of non-linear combinations of the Ricci tensor is expected to  contain ghosts \cite{1978GReGr...9..353S,2005JCAP...03..008C} when expanded around a Minkowski or vacuum background (except when the scalars enter in the Gauss-Bonnet combination). However, a separate discussion of the existence of ghosts or other instabilities  is left to future work, also because models based on the Ricci-inverse might even lack a Minkowskian limit.

Finally, we conclude this work  by  pointing out some possibilities to escape the no-go theorem.

\section{Main equations}
Consider first the basic  Action 
\begin{equation}
S=\int\sqrt{-g}d^{4}x (R+\alpha A)\label{eq:first} \; ,
\end{equation}
where the anticurvature scalar $A$ is the trace of $A^{\mu\nu}$
\begin{equation}
A^{\mu\nu}=R_{\mu\nu}^{-1} \; .
\end{equation}
 By differentiating Eq. (\ref{eq:delta}), we see that 
\begin{equation}
\delta A^{\mu\tau}  =-A^{\mu\nu}(\delta R_{\nu\sigma})A{}^{\sigma\tau} \; .
\end{equation}
  We have then 
\begin{align}
\delta S & =\int d^{4}x(A\delta\sqrt{-g}+\sqrt{-g}A^{\mu\nu}\delta g_{\mu\nu}+\sqrt{-g}g_{\mu\nu}\delta A^{\mu\nu})\\
 & =\int d^{4}x\sqrt{-g}(\frac{1}{2}Ag^{\mu\nu}\delta g_{\mu\nu}+A^{\mu\nu}\delta g_{\mu\nu}+g_{\mu\nu}\delta A^{\mu\nu}) \; ,
\end{align}
and since
\begin{equation}
\delta R_{\alpha\beta}=\nabla_{\rho}\delta\Gamma_{\beta\alpha}^{\rho}-\nabla_{\beta}\delta\Gamma_{\rho\alpha}^{\rho} \; ,
\end{equation}
we obtain 
\begin{align}
\delta A^{\mu\nu} & =-A^{\mu\alpha}(\nabla_{\rho}\delta\Gamma_{\beta\alpha}^{\rho}-\nabla_{\beta}\delta\Gamma_{\rho\alpha}^{\rho})A^{\beta\nu}\\
 & =-\frac{1}{2}A^{\mu\alpha}(g^{\rho\lambda}\nabla_{\rho}(\nabla_{\alpha}\delta g_{\beta\lambda}+\nabla_{\beta}\delta g_{\lambda\alpha}-\nabla_{\lambda}\delta g_{\alpha\beta})-g^{\rho\lambda}\nabla_{\beta}(\nabla_{\alpha}\delta g_{\rho\lambda}+\nabla_{\rho}\delta g_{\lambda\alpha}-\nabla_{\lambda}\delta g_{\alpha\rho}))A^{\beta\nu}\\
 & =-\frac{1}{2}A^{\mu\alpha}g^{\rho\lambda}(\nabla_{\rho}\nabla_{\alpha}\delta g_{\beta\lambda}-\nabla_{\rho}\nabla_{\lambda}\delta g_{\alpha\beta}-\nabla_{\beta}\nabla_{\alpha}\delta g_{\rho\lambda}+\nabla_{\beta}\nabla_{\lambda}\delta g_{\alpha\rho} + \left[\nabla_{\beta},\nabla_{\rho} \right]\delta g_{\lambda\alpha})A^{\beta\nu} \; .
\end{align}
Using integration by parts, this becomes
\begin{align}
g_{\mu\nu}\delta A^{\mu\nu} & =-\frac{1}{2}g_{\mu\nu}g^{\rho\lambda}(\delta g_{\beta\lambda}\nabla_{\alpha}\nabla_{\rho}(A^{\mu\alpha}A^{\beta\nu})-\delta g_{\alpha\beta}\nabla_{\lambda}\nabla_{\rho}(A^{\mu\alpha}A^{\beta\nu})-\delta g_{\rho\lambda}\nabla_{\alpha}\nabla_{\beta}(A^{\mu\alpha}A^{\beta\nu})+\delta g_{\alpha\rho}\nabla_{\lambda}\nabla_{\beta}(A^{\mu\alpha}A^{\beta\nu})) \\ 
 & =\frac{1}{2}\delta g_{\iota\kappa}(-2g^{\rho\iota}\nabla_{\alpha}\nabla_{\rho}A^{\mu\alpha}A_{\mu}^{\kappa}+\nabla^{2}(A^{\mu\iota}A_{\mu}^{\kappa})+g^{\iota\kappa}\nabla_{\alpha}\nabla_{\beta}(A^{\mu\alpha}A_{\mu}^{\beta})) \; .
\end{align}
So finally the variation is 
\begin{align}
\delta g_{\mu\nu}(\frac{1}{2}Ag^{\mu\nu}+A^{\mu\nu}+\frac{1}{2}(-2g^{\rho\mu}\nabla_{\alpha}\nabla_{\rho}A^{\sigma\alpha}A_{\sigma}^{\nu}+\nabla^{2}(A^{\sigma\mu}A_{\sigma}^{\nu})+g^{\mu\nu}\nabla_{\alpha}\nabla_{\beta}(A^{\sigma\alpha}A_{\sigma}^{\beta}))) \; .
\end{align}
Together with the variation of the standard Hilbert-Einstein Lagrangian  
\begin{align}
\delta g^{\mu\nu}(-\frac{1}{2}Rg_{\mu\nu}+R_{\mu\nu}) & =-\delta g_{\mu\nu}(-\frac{1}{2}Rg^{\mu\nu}+R^{\mu\nu}) \; ,
\end{align}
we obtain finally the equations for the Action (\ref{eq:first})
\begin{align}
R^{\mu\nu}-\frac{1}{2}Rg^{\mu\nu}-\alpha A^{\mu\nu}-\frac{1}{2}\alpha Ag^{\mu\nu}+\frac{\alpha}{2}\left(2g^{\rho\mu}\nabla_{\alpha}\nabla_{\rho}A_{\sigma}^{\alpha}A^{\nu\sigma}-\nabla^{2}A_{\sigma}^{\mu}A^{\nu\sigma}-g^{\mu\nu}\nabla_{\alpha}\nabla_{\rho}A_{\sigma}^{\alpha}A^{\rho\sigma}\right) & =T^{\mu\nu}\label{eq:eom1} \; ,
\end{align}
where we used the fact that $A^{\alpha}_{\sigma }A^{\nu\sigma}=A^{\alpha\tau}g_{\tau\sigma}A^{\sigma\nu}=A^{\alpha\tau}A_{\tau}^{\nu}=A^{\alpha\sigma}A_{\sigma}^{\nu}=A_{\sigma}^{\nu}A^{\alpha\sigma}$ and we employed units in which $8\pi G=1$. It can be show that the left-hand side of Eq.~\eqref{eq:eom1} is divergenceless, as it should be in order to satisfy the Bianchi identities.

The extension to any Lagrangian $f(R,A)$ is quite straightforward: 
\begin{equation}
\delta S=\int d^{4}x\sqrt{-g}(-\frac{1}{2}f(R,A)g_{\mu\nu}\delta g^{\mu\nu}+f_{A}A^{\mu\nu}\delta g_{\mu\nu}+f_{A}g_{\mu\nu}\delta A^{\mu\nu}+f_{R}R_{\mu\nu}\delta g^{\mu\nu}+f_{R}g^{\mu\nu}\delta R_{\mu\nu}) \; ,
\end{equation}
where $f_R=\partial f/\partial R$ and $f_A=\partial f/\partial A$. Then we have
\begin{align}
f_{R}R^{\mu\nu}-f_{A}A^{\mu\nu}&-\frac{1}{2}fg^{\mu\nu}+g^{\rho\mu}\nabla_{\alpha}\nabla_{\rho}f_{A}A^{\alpha}_{\sigma}A^{\nu\sigma}-\frac{1}{2}\nabla^{2}(f_{A}A_{\sigma}^{\mu}A^{\nu\sigma}) &\nonumber\\&-\frac{1}{2}g^{\mu\nu}\nabla_{\alpha}\nabla_{\beta}(f_{A}A_{\sigma}^{\alpha}A^{\beta\sigma})-\nabla^{\mu}\nabla^{\nu}f_{R}+g^{\mu\nu}\nabla^{2}f_{R}  =T^{\mu\nu} \; .\label{eq:master}
\end{align}
It is well known that one can recast a $f(R)$ theory in the form of a scalar-tensor theory in the Einstein frame introducing a scalar field non minimally coupled to gravity. 
Usually this is done by defining a scalar field $\phi = df/dR$ and performing a Legendre transformation of the function $f$. Such an approach, however, 
fails here, because $A$ is not a one-to-one function of $R$ and therefore
 $df(A)/dR$ is in general not invertible. In Ref. \cite{2005JCAP...03..008C}
 a general way to recast Lagrangians based on functions of the Ricci tensor as multi-scalar-tensor theories is discussed, but this approach is not particularly helpful so it will not be pursued here.
 
A code that evaluates the equations of motion for any $f(R,A)$ in a given metric is made publicly available.\footnote{https://github.com/itpamendola/inverse-ricci}

We note in passing that on a deSitter cosmological solution all the terms with  derivatives in Eq. (\ref{eq:master}) vanish. Taking the trace one has then in vacuum
\begin{align}
f_{R}R-f_{A}A-2f =0 \; ,
\end{align}
Since on deSitter $R=12H^2$ and $A=4/(3H^2)$, this equation can be easily solved for any $f(R,A)$ model to check whether one gets non-trivial (i.e. $H\not = 0$) solutions that could replace a cosmological constant. For instance, if $f=R-\alpha A$ (where $\alpha$ is a constant with dimensions $H_0^4$) then we see that $H=(\alpha/3)^{1/4}=const$. 

\section{A no-go theorem}

Let us now write down the anticurvature scalar $A$ in a flat-space FLRW metric. It is easily found that
\begin{equation}
    A=\frac{2(6+5\xi)}{3H^2(1+\xi)(3+\xi)} \; ,
\end{equation}
where $\xi\equiv H'/H$ and a prime stands for $d/d\log a$. Notice that although $A$ is singular in the Minkowski limit ($H\to 0$), as obviously expected,  $A^{-1}$ is not.
We see however that $A$ is singular\footnote{Unless at the same time $H$ suitably vanishes or diverges: this would however only occur for special initial conditions.}  for $\xi=-3,-1$ and vanishes for $\xi=-6/5$.
This means that if the cosmic evolution passes through any one of these values of $\xi$, either $A$ or $A^{-1}$, or any of their powers, develops a singularity. If during the evolution $A$ passes through {\it both} 0 and $\pm \infty$, then any term in the Lagrangian that contains $A^n$, for $n$ positive or negative, will blow up. This behavior will reflect into equations of motion that also contain a singularity at the same cosmic epochs. Now we show that this is exactly what happens.

The quantity $\xi$ can also be written as
\begin{equation}
    \xi=-\frac{3}{2} (1+w_{\rm eff}) \; ,
\end{equation}
where $w_{\rm eff}$ is the total equation of state, and (just like $H$ or $\xi$) is what current distance observations  measure (through the integral $\int dz/H$ that appears in the expression for the cosmic distance). In a $\Lambda$CDM model, $w_{\rm eff}=w_\Lambda \Omega_\Lambda=-\Omega_\Lambda$.
Now,  observations \cite{Abbott:2018wog,Scolnic:2017caz,Aghanim:2018eyx,Ade:2015xua} tell us that the Universe evolved from a decelerated phase with $w_{\rm eff}\approx 0$ (so $\xi\approx -1.5$) into an accelerated phase $w_{\rm eff}\approx -0.7$ (so $\xi\approx -0.45$). Therefore the cosmic expansion had to pass, at redshifts around unity, through both $\xi=-1$ and $\xi=-6/5$. This demonstrates that $A$ and $A^{-1}$ will both be singular at some epoch between deceleration and acceleration.
Consequently, any Lagrangian that contains additive terms proportional to $A^n$ (e.g. the two simplest scale-free models, $f(R,A)=R+\alpha A^{-1}$ and $f(R,A)=R+\alpha R^2 A$, with $\alpha$ a dimensionless constant) are ruled out as dark energy models. Notice also that $R=6H^2( \xi+2)$ so no power of $R$ can cure the singularity.

Before speculating on how to avoid this problem, let us show its realization in the simple case $f=R+\alpha A^{-1}$.

\section{Lagrangian $R+\alpha/A$}

In this case we find the modified Friedmann equation 
\begin{equation}
 \rho_t= 3 \alpha  H^2\frac{ (\xi +3)^2 (5 \xi +6)-18 \xi '}{4 (5 \xi +6)^3}+3 H^2 \; ,
\end{equation}
and the $(i,i)$ equation
\begin{align}
w_t\rho_t=-\frac{\alpha  H^2 \left[(5 \xi +6) \left((\xi +3)^2 (2 \xi +3) (5 \xi +6)-18 \xi ''\right)+270 \left(\xi '\right)^2-54
   (\xi +2) (5 \xi +6) \xi '\right]}{4 (5 \xi +6)^4}-2 H^2 \xi -3 H^2 \; ,
\end{align}
(here the subscript $t$ designs total matter)
plus of course the matter conservation equation $\rho_t'=-3(1+w_t)\rho_t$.  
The expected singularity at $\xi=-6/5$ appears in both equations. The $A$ energy density is given by
\begin{equation}
  \Omega_A\equiv  -\alpha\frac{    (\xi +3)^2 (5 \xi +6)-18 \xi '}{4 (5 \xi +6)^3} \; .
\end{equation}
 
 \begin{figure}
\includegraphics[width=10cm]{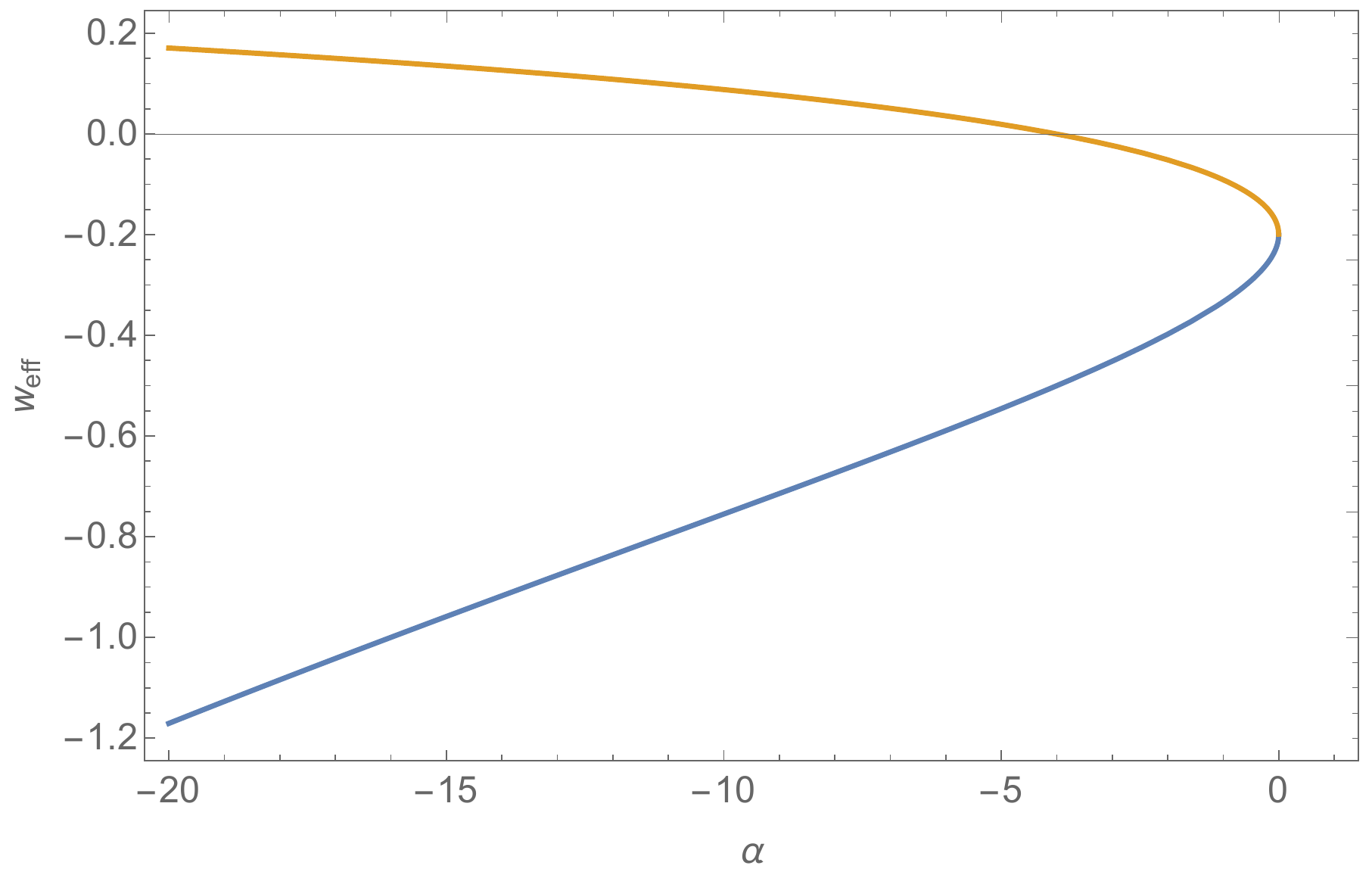}
\caption{The two real branches of the function $w_{\rm eff}(\alpha)=-2\xi/3-1$ from Eq. (\ref{eq:walpha})\label{fig:walpha}}
\end{figure}

We consider now two cases, first with just pressureless matter, then adding a cosmological constant.
In the first case, $w_t=0$ and we obtain a single equation for $\xi$: 
\begin{equation}
   \xi''= \frac{15 \left(\xi '\right)^2}{5 \xi +6}-3 (\xi +2) \xi '+\frac{(2 \xi +3) (5 \xi +6) \left(\alpha  (\xi +3)^2+4 (5 \xi +6)^2\right)}{18
   \alpha } \; .
\end{equation}

We can now study qualitatively the system by searching for critical points and determining their stability. We find critical  points for $\xi=const$, which also give $\Omega_m=1-\Omega_A=const$. These solutions correspond to power-law expansions with scale factor $a\sim t^n$, where $w_{\rm eff}=-1+3n/2$. We find critical points at 
$\Omega_m= 1+\frac{\alpha}{4}$ and $\xi_0=-\frac{3}{2}$, and for
\begin{equation}
 \Omega_m=0\,,\quad   \xi_{\pm}=\frac{3(-40-\alpha\pm 6 \sqrt{-\alpha})}{100+\alpha} \; .
 \label{eq:walpha}
\end{equation}
In Fig. \ref{fig:walpha} we illustrate $\xi_{\pm}(\alpha)$. We see that for every $\alpha<0$ there are two  real solutions, one above, the other below $\xi=-6/5$, or equivalently $w_{\rm eff}=-0.2$. Some of these solutions seem  cosmologically interesting. For instance, for $\alpha\approx -8$, the two solutions correspond to the observed present accelerated value $w_{\rm eff}\approx -0.67$ 
and to an expansion quite close to a matter dominated era,  $w_{\rm eff}\approx 0.06$. Analogously,
if $\alpha=-4$, one has $w_{\rm eff}=0$, i.e. an exact matter era evolution without matter, in which the $A$ energy density acts as a form of dark matter. The other solution, $\xi_+$, corresponds to $w_{\rm eff}=-0.5$, i.e. an accelerated solution still marginally compatible with observations. A cosmic evolution that moves from one such solution to the other would be indeed an intriguing possibility, replacing both dark matter and dark energy with the anticurvature tensor without any new scale nor fine-tuned parameters. However, as anticipated, this does not occur.

\begin{figure}[tb]
    \centering
    \includegraphics[width=10cm]{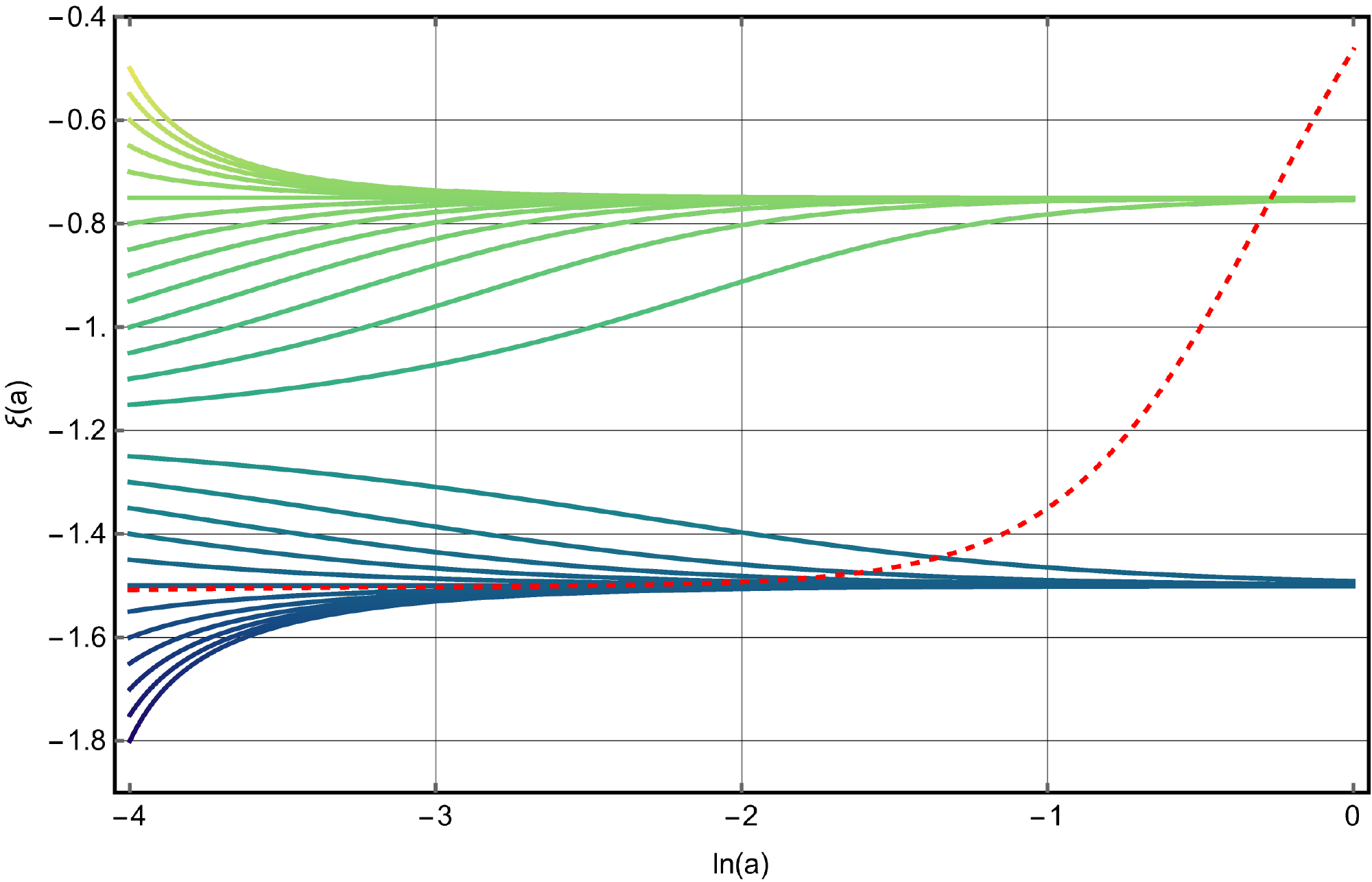}
    \caption{ Numerical solutions $\xi(a)$ of Eq. (25) in case of $\Omega_m \neq 0$, $\Omega_\Lambda=0$ with $w=0$ and $\alpha=-4$. The  solutions $\xi=-3/2$ and $\xi=-3/4$ are confirmed to be attractors. The divide at $\xi=-3/2$ is also evident. The red dashed line is the $\Lambda$CDM behaviour.}
    \label{fig:ximatter}
\end{figure}

Through a stability analysis of the linearized dynamical system we find that the the critical point $\xi_{-}$ is a stable attractor only for $-4\le\alpha\le 0$.
The critical point $\xi_+$ is a stable attractor for $\alpha \leq 0$, while the linear analysis alone cannot assess the stability of the point $\xi = -3/2$. These findings are supported by the numerical investigation shown in Fig. \ref{fig:ximatter}, so that the cosmic evolution will end up either at $\xi_+$ or $\xi_{-}$, depending on whether the initial $w_{\rm eff}$ is above or below the singularity at  $w_{\rm eff}=-0.2$. The crucial point is that no trajectory can cross  the $w_{\rm eff}=-0.2$ ridge; consequently, as anticipated on general grounds, the cosmic expansion cannot move from a decelerated phase around $w_{\rm eff}=0$ to an accelerated one around $w_{\rm eff}\approx -0.7$.

\begin{figure}[tb]
    \centering
    \includegraphics[width=10cm]{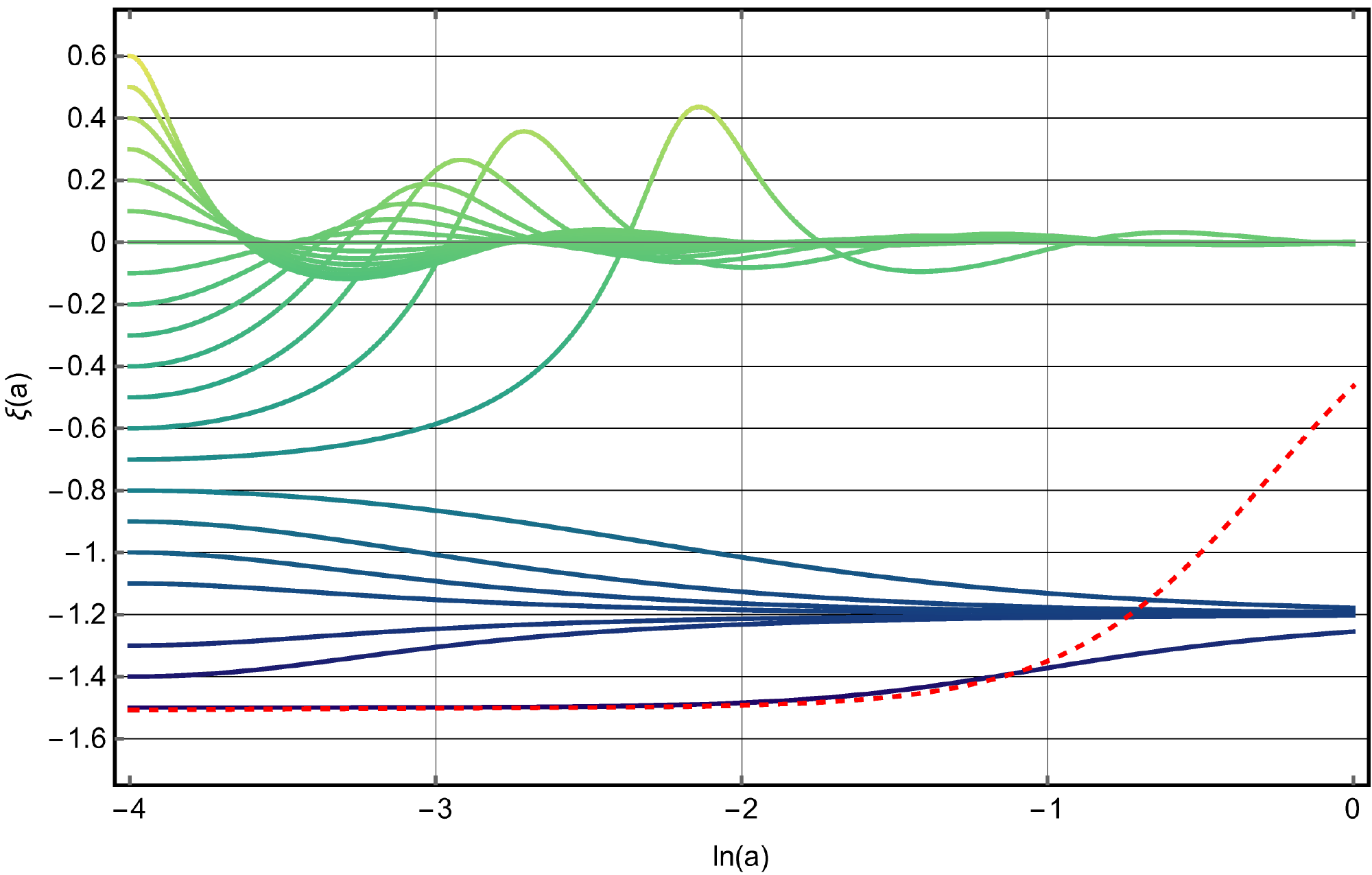}
    \caption{ Numerical solutions $\xi(a)$ of Eq.~\eqref{eq:mattercosmconst} with matter and cosmological constant, for $\alpha=-4$. The upper curves converge toward the de Sitter attractor at $\xi=0$. The lower curves converge towards  the divide line at $\xi=-6/5$, which is now also an attractor. The red dashed line is the $\Lambda$CDM behavior.}
    \label{fig:ximatterlambda}
\end{figure}

When matter is composed of dust plus a cosmological constant, we derive instead the following  equation for $\xi,\Omega_m$,
\begin{equation}
\xi ''= \frac{6 (5 \xi +6)^4 \Omega _m+\xi  (5 \xi +6)^2 \left(9 (\alpha +16)+(\alpha +100) \xi ^2+6 (\alpha +40) \xi \right)+135
   \alpha  \left(\xi '\right)^2-27 \alpha  \left(5 \xi ^2+11 \xi +6\right) \xi '}{9 \alpha  (5 \xi +6)} \; ,
   \label{eq:mattercosmconst}
\end{equation}
to be complemented by the conservation equation
\begin{equation}
\Omega_{m}'=-(3+2\xi)\Omega_{m} \; .
\end{equation}
The phase space now is more complicated, in particular we found that now $\xi_-$ is always unstable and $\xi_+$ is a stable attractor for $\alpha <-16$. The critical point $\xi = -3/2$ is always unstable, while a new critical point $\xi = 0$, i.e. a de Sitter state, appears  which is a stable attractor when $-16 \leq\alpha \leq0$. However, the bottom line is the same, as can be immediately gleaned from Fig. \ref{fig:ximatterlambda},  so the model is ruled out as a candidate for dark energy even when a cosmological constant is added, regardless of the value of $\alpha$.

\begin{figure}[tb]
    \centering
    \includegraphics[width=10cm]{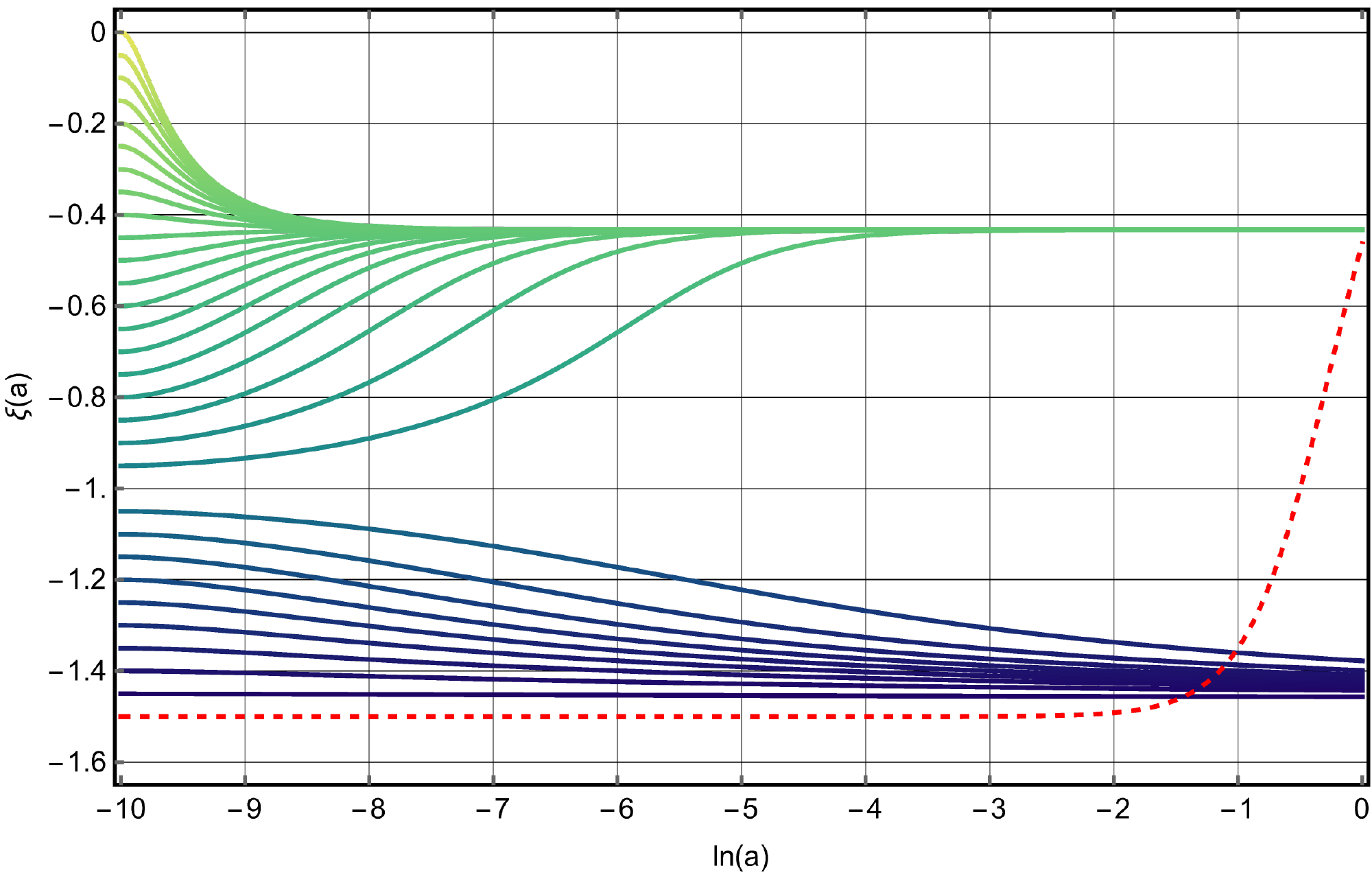}
    \caption{ Numerical solutions $\xi(a)$ for the Lagrangian $L=R+\alpha R^2 A$ in the presence of matter for $\alpha=-0.4$. The upper curves converge toward the accelerated attractor at $\xi=-0.43$. The lower curves converge towards the decelerated attractor at $\xi=-1.46$. The divide at $\xi=-1$ is evident. The red dashed line is the $\Lambda$CDM behavior.}
    \label{fig:ximatterR2}
\end{figure}

We also analysed numerically the case $L=R+\alpha R^2 A$ and found a qualitatively very similar behavior (see Fig. \ref{fig:ximatterR2}), now with a divide at $\xi=-1$ as expected.

\section{Circumventing the no-go theorem}

Here we discuss some ways to avoid the no-go theorem.

\subsection{Spatially curved metric}
The first possibility is to move away from a flat-space FLRW.
 If there is a non-zero spatial curvature then
\begin{equation}
    A=\frac{2(6+5\xi+3\Omega_k)}{3H^2(1+\xi)(3+\xi+6\Omega_k)} \; ,
\end{equation}
and the singularity of $A^{-1}$ is shifted to another value of $\xi$, while the one of $A$ at $\xi=-1$ is not modified. However, if $\Omega_k$ is very close to zero, as observations show \cite{Aghanim:2018eyx}, then the shift will be very small, and the singularity  remains between the decelerated and accelerated phases.

\subsection{Anisotropic background}

Another simple possibility is to consider an anisotropic universe. Let us adopt for illustrative purpose the simplest choice, a Bianchi I spacetime:
\begin{equation}\label{BImetric}
    ds^2 = -dt^2 +a(t)^2\left(e^{2\beta_x(t)}dx^2 +e^{2\beta_y(t)}dy^2 + e^{2\beta_z(t)}dz^2 \right) \; ,
\end{equation}
where we have defined the averaged scale factor
\begin{equation}
    a(t) = \sqrt[3]{a_x(t) a_y(t) a_z(t)}\; ,
\end{equation}
so that $a_i(t) = a(t)e^{\beta_i}$, and the $\beta_i$ satisfies $\sum_i \beta_i = 0$.
For the sake of simplicity, let us specialise to the case $\beta_x = -\beta_z \equiv \beta$ and $\beta_y = 0$. In this case the anticurvature scalar $A$ reads:
\begin{equation}
    A = \frac{1}{H^2}\left[\frac{4\xi + 6 + \frac{(\beta')^2}{2}}{3\left(3+\xi\right)\left(1+\xi + \frac{(\beta')^2}{6}\right)} + \frac{2(3+\xi)}{\left(3+\xi\right)^2 - \frac{1}{4}\left(\beta'' + \beta'(3+\xi)\right)^2}  \right] \; ,
\end{equation}
which for $\beta'=0$ reduces to the FLRW case.
As we can see, the singularity $\xi = -1$ is shifted by the anisotropic term $(\beta')^2/6$. Note  that also the singularity appearing in $A^{-1}$, $\xi = -6/5$, is in general shifted. For example, if $\beta''$ is negligible, we have that $A^{-1}$ is singular for
\begin{equation}
    \xi = \frac{24-\frac{\beta'^4}{4}}{ 2\beta'^2-20}\approx -\frac{6}{5}(1-\frac{(\beta')^2}{10})\;,\label{betasingular}
\end{equation}
(the last approximate equality being valid for  $\beta'\ll 1$) which recovers the FLRW case for $\beta'=0$, while being regular in $\xi= -6/5$ unless $\beta'^2 = 48/5$, i.e. the two roots of Eq.~\eqref{betasingular} for $\xi = -6/5$. 
This shows that relaxing the assumption of spatial isotropy the singularities occurring in the anticurvature scalar and its inverse can be arbitrarily shifted, but not removed. It is clear however that one needs $\beta'$ of order unity to move the singularity outside the observational range, which on the other hand is not likely to be compatible with experimental data.

To illustrate that, let us naively estimate $\beta$ from the evidence of anisotropic expansion claimed recently in \cite{Migkas:2020fza}, emerged from X-ray observations of galaxy clusters. Here the authors find that the highest and the lowest values observed for the universe expansion rate are $H_{\rm max} \sim 75 \;$  km/s/Mpc and $H_{\rm min} \sim 66 \;$  km/s/Mpc. Identifying $H_{\rm max}$ with $H_x = H + H\beta'$ and $H_{\rm min}$ with $H_z = H -H\beta'$ it is easy to compute:
\begin{equation}
    H_{\rm max}-H_{\rm min} = H_x - H_z = 2 H \beta' \sim 9 \;  {\rm km/s/Mpc} \; ,
\end{equation}
from which, assuming that the averaged Hubble factor is $H \sim 70 \;$ km/s/Mpc, we obtain
\begin{equation}
    \beta' \sim 0.06 \; ,
\end{equation}
which shows that generally $\beta'$ is constrained from the observations to be too small to shift the singularities of $A$ outside the observational range.

\subsection{Revising the data analysis}

The no-go theorem is based on the distance observations through e.g. supernovae Ia (other data, like CMB, are more model-dependent and a completely new analysis based on the present theory should be performed). These distance observations measure directly the luminosity or the angular diameter distances and only indirectly derive $H$. One could then imagine some contrived model in which the past expansion is compatible with $w_{\rm eff}<-0.2$, such that the final acceleration can be reached without passing through the  $w_{\rm eff}= -0.2$ ridge. However, one should also find a way to produce a viable radiation era, when $w_{\rm eff}=1/3$, so this way out seems very unlikely.

\subsection{Non-polynomial Lagrangians}

The previous 
attempts at finding loopholes in the no-go theorem are not satisfactory. The last possibility we briefly discuss is to design a
 Lagrangian function of $A$ that remains regular both for $A\to 0$ and $A\to\pm\infty$. In this case, in fact, the singularities of the Lagrangian are avoided. For instance, still considering  scale-free Lagrangians for simplicity, $R+\alpha R\exp[-\beta (RA)^2]$ or $R/(1+\alpha RA)$ have this property. In the former case, for example, the Friedmann equation around the critical points, i.e. assuming $\xi'=\xi''=0$, becomes:
 \begin{equation}
     3 H^2 \left(1-\frac{\alpha \mathcal{P}_5(\xi,\beta)}{(\xi +1)^3 (\xi +3)^2}e^{-\frac{16 \beta(\xi +2)^2 (5 \xi +6,)^2}{(\xi +1)^2 (\xi +3)^2}}\right) = \rho_m \; ,
 \end{equation}
 where $\mathcal{P}_5(\xi,\beta)$ is a polynomial of order five in $\xi$ and linear in $\beta$. It is straightforward to realise that the above equation is regular on the poles of the denominator due to the presence of the exponential factor. One could then work out the cosmological implications of such models, but this is beyond the scope of this work. 
 
Another option is to include scalar combinations of higher order in the anticurvature tensor, like $A^{\mu\nu}A_{\mu\nu}$. In FLRW background the latter looks as follows:
\begin{equation}
    A^{\mu\nu}A_{\mu\nu} = \frac{4}{9H^4}\frac{7\xi^2 + 15\xi + 9}{(\xi+1)^2(\xi + 3)^2} \; .
\end{equation}
We see that it still contains the singularities at $\xi = -3$ and $\xi = -1$, but remarkably it never vanishes. This means that a theory from a Lagrangian $(A^{\mu\nu}A_{\mu\nu})^{-1}$ should be free of this kind of singularities.

A similar approach would be to assume that the $A$ terms are only effective at some early cosmic epoch, e.g. during inflation. In this case, the no-go theorem would also be irrelevant and the anticurvature theory could introduce some interesting phenomenology.

\section{Conclusions}

The anticurvature tensor, defined as the inverse of the Ricci tensor, allows the formulation of an alternative theory of gravity. Such a theory will have to be analysed to assess the existence of ghosts or other instabilities and derive the effects on perturbations such as the propagation of gravitational waves,  the growth of perturbations, or the Newtonian limit. After deriving the general equations of motion for a generic Lagrangian function of the curvature and the anticurvature scalars, here we concerned ourselves about whether such a theory could be a viable candidate as a cosmological model. We have shown that a general no-go theorem prevents cosmic trajectories to join a decelerated phase to an accelerated one in any Lagrangian that contains a term with any positive or negative power of the anticurvature scalar, thereby ruling out a vast class of models. This no-go theorem is illustrated analytically and numerically  in some particularly simple cases of models without new dimensional scales. We discuss possible ways out of the theorem, and in particular we point out some (more complicate) Lagrangians that avoid it. We leave to future work a systematic study of such models, in search of novel phenomenology.

\section*{Acknowledgments}
We thank Tomi Koivisto, Ignacy Sawicki, and Oliver Piattella   for useful comments on the draft.
L.G. is  grateful to the ITP-Heidelberg for the kind hospitality. His  work was financed in part by the Coordena\c{c}\~ao de Aperfei\c{c}oamento de Pessoal de N\'{i}vel Superior - Brasil (CAPES) - Finance Code 001 and from the DAAD Co-financed Short-Term Research Grant Brazil, 2019 (57479964).
L.A. acknowledges  support from DAAD PPP Brasilien 57518956 (2020).

\bibliographystyle{apsrev4-2.bst}
\bibliography{AR.bib}
\end{document}